\documentclass[referee]{raa} 
\usepackage{lscape}           
\usepackage{natbib}

\bibstyle{model2-names.bst}
\usepackage{graphicx,times}             
\usepackage{amssymb,amsmath}
\bibpunct{(}{)}{;}{a}{}{,}

\usepackage[a4paper=true,dvipdfm=true,pagebackref=true]{hyperref}
\hypersetup{colorlinks = true, linkcolor = green, anchorcolor = red, citecolor = blue, filecolor = red, pagecolor = red, urlcolor = red}

\begin{document}

   \title{Testing the Effect of  Solar Wind Parameters and Geomagnetic  Storm Indices  on  Galactic Cosmic Ray Flux Variation with Automated-Selected Forbush Decreases
}

   \volnopage{Vol.0 (20xx) No.0, 000--000}      
   \setcounter{page}{1}          

   \author{J. A. Alhassan
      \inst{1*}
   \and O. Okike
      \inst{2}
   \and A. E. Chukwude
      \inst{1}
   }

   \institute{Department of Physics and Astronomy, University of Nigeria, Nsukka, Nigeria \\
        \and
             Department of Industrial Physics, Ebonyi State University, Abakaliki, Nigeria\\
{\it *Corr. Author:jibrin.alhassan@unn.edu.ng}\\ 
\vs\no
   {\small Received~~20xx month day; accepted~~20xx~~month day}}

\abstract{ Forbush decrease (FD), discovered by Scott E. Forbush about 80 years ago, is  reffered to  as the non-repetitive short-term depression in galactic cosmic ray (GCR) flux, presumed to be associated with large-scale perturbations in solar wind  and interplanetary magnetic field (IMF). It is  the most spectacular variability in the GCR intensity which  appear to be the compass for investigators seeking  solar-terrestrial relationships. The method of selection and validation of FD events are very important to cosmic ray scientists. We have deployed a  new computer software   to determine the amplitude and timing of FDs from daily-averaged cosmic ray (CR) data at OULU neutron monitor station. The code selected 230 FDs between 1998 and 2002. In an attempt to validate the new  FD automated  catalog, the relationship between the amplitude of FDs, and IMF, solar wind speed (SWS) and geomagnetic storm indices (Dst, kp, ap) is tested here. A  two-dimensional regression analysis indicates  significant  linear relationship  between large FDs (CR(\%) $\leq-3$) and solar wind data and geomagnetic storm indices in the present sample. The implications of the relationship among these parameters are discussed.
}

   \authorrunning{J. A. Alhassan, O. Okike \& A.E Chukwude } 
   \titlerunning{ Effect of  Solar wind Parameters and Geomagnetic  Storm  indices}  

   \maketitle

%
\section{Introduction}           
\label{sect:intro}

Forbush decrease (FD), discovered by Scott E. Forbush  \citep{fo:37, fo:38}, is one of the outstanding  transient changes in cosmic-ray flux, observed by
ground-based neutron detectors \citep{bad:2015}. It is referred to as the non-repetitive short-term depression in galactic cosmic ray (GCR) intensity  presumed to be associated with large-scale perturbations in solar wind  and interplanetary magnetic field (IMF). Generally, FDs can be classified as recurrent and non-recurrent in accordance with their solar sources. Whereas the recurrent FDs are induced by high-speed solar wind streams (HSSWs)  from coronal holes (CHs) which rotate together with the sun, the non-recurrent FDs  are triggered by coronal mass ejections (CMEs) and their  interplanetary extensions (ICMEs) \citep{richard:2004, richard:2011, be:2014}. Forbush events from HSSWs are characterized by small amplitude, gradual onset and symmetric profile while FDs caused by ICMEs have signatures of large magnitude, sudden onset and gradual recovery  \citep{ Lockwood:1971,avbelov:2009,Melkumyan:2019}. 

Geomagnetic storms (GMSs) which is  a temporary disturbance of the Earth's magnetosphere and non-recurrent Forbush events  are both presumed to have a common solar and interplanetary origin in the form of CMEs and ICMEs \citep[e.g.][]{richardson:2011, chertok:2015, bad:2015}. ICMEs are conveyed from the Sun to the Earth by a stream of  charged particles known as solar wind (SW). In the course of their transportation to the Earth, CMEs and ICMEs interact with galactic cosmic rays that fill  the interplanetary space to generate interplanetary magnetic filed (IMF) and shock waves \citep[e.g.][]{richardson:2011, lin:2016, ok:2020e}. The IMF disturbances have been found to be associated with FD \citep{oh:08,oh:09}. Investigations of cosmic-ray intensity variations have been carried out with data recorded by ground-based and space-borne equipment for over five decades now \citep{fan:60, be:08}. Yet, association of FDs  with geomagnetic activity indices and solar wind disturbances is a subject of continuous interest  \citep[e.g.][]{be:05, be:08, lin:2016, chris:20}. Understandably,  disturbances in  solar wind, magnetosphere and cosmic rays are caused by the same active processes in the Sun and thus  interrelated, hence the rationale for investigating them together in this work. 

It has been argued that FD is  the most spectacular variability in the GCR intensity which also appear to be the compass for investigators seeking for solar terrestrial relationships \citep{ok:2020}. Understanding of Sun\textendash  Earth weather connections require FD as one of the mediator parameters to be employed.  Obtaining large dataset of FD event is very critical before any statistically reliable investigation can be carried out. Publications emanating from IZMIRAN group have  investigated large FDs based on semi-automated global  survey method (GSM) \citep[e.g.][]{avbelov:2001, avbelov:2009, be:2014, be:2015}. However, other  investigators select  few FD catalogues with the  manual technique for their research \citep[e.g.][]{oh:08, kris:08, to:2001, to:2004, calo:2010, sven:2016}. FD being a key parameter in space weather task and in addition to raw CR data inherent many signal superposition tendencies,  its selection   requires sophisticated computer technique rather than the  manual method \citep{rami:2013}. Validation of FDs, selected by the manual, semi-automated or fully automated approach is not a very common practice among CR scientists.

In this paper, we intend to validate the algorithm selected FD list from Oulu neutron monitor station between  1998 to 2002  by testing the correlation between its amplitudes,  solar wind data and geomagnetic activity indices. 

\section{Data}
We have taken Oulu neutron monitor (NM) pressure-corrected raw cosmic ray intensity daily averaged data between 1998 to 2002  from IZMIRAN common website : \url{http://
cr0.izmiran.ru/common} for our investigation in this work. The site is hosted by the Pushkov Institute of Terrestrial Magnetism, Ionosphere, and Radio Wave Propagation, Russian Academy of Sciences (IZMIRAN) team. Oulu NM station is characterized by geomagnetic cutoff rigidity of 0.77 GV, geographic location of  25.47$^{\circ}$E, 65.05$^{\circ}$N and altitude = 0 m. Daily mean solar wind speed (SWS), IMF, disturbance storm-time index (Dst), geomagnetic indices kp and ap  data are obtained from the OMNI database https:
//omniweb.gsfc.nasa.gov/html/ow data.html.

\section{Manual FD detection technique}

CR data in  raw form are fraught with  superposition
of several signals of similar or different magnitudes, cycles, and periodicities \citep{ok:2019c}. Timing and amplitude estimation of  true FDs will not be easy with the traditional manual method. In the manual technique, the time domain component is not factored into the calculation as shown in equation 1 \citep[e.g.][]{har:2010, lee:2013, oki_:2020, okike:2020c}.

\begin{equation} \label{eqn}
FD_ {mag} (\%) = \frac {FD_{min+1}-FD_{min-1}}{FD_{min}}\times 100\% ,
\end{equation}\\
where $FD_{mag} (\%)$ is the FD magnitude, $FD_{min}$ represent CR count on the day of minimum depression, $FD_{min+1}$, the CR count rate a day before minimum depression, and $FD_{min-1}$, CR count rate a day after minimum depression.\\

The Manual method   involves several steps. It includes  downloading of CR data of prefered resolution, plotting and visually searching for pits or points of minimum depressions in CR data, identifying the onset  and endtime of CR counts from the main phase profile and finally calculating the individual FD amplitude. This technique is very tasking, subjective and time consuming though employed  by many  investigators over the years \citep[e.g.][]{oh:08, kris:08, oh:09, le:2015}. The absence of time domain factor is a major demerit in the manual FD selection technique. CR data being a single time-dependent series signal can only be handled by computer software.
 
\section{Automated FD Location Code}
The  automated FD location code in this work is a modified version of event selection program developed by \citet{ok:2019}. This technique takes  unprocessed CR data as its input signal instead of the high frequency signal. It is   designed to simultaneously search for depressions/turning points and the time of the reductions in the raw CR data. While \citet{ok:2019} identification code requires several subroutines that perform a number of tasks such as Fourier transformation, data filtering, and event timing, the current code is simpler. A few subprograms which computes event amplitude and timing are involved. In the manual technique \citep[see][for example]{le:2015, har:2010, okike:2020c}, equation \ref{eqn} is used to calculate individual CR percentage (with reference to hourly, daily, monthly or even annual averages) reduction, whereas the equation is used for normalization of CR time series in \citet{ok:2019} and in the current work. 

The depressions in the normalized data are indications of FD events. The two subroutines (timing and magnitude estimation routines) incorporated in the program are used to search for the depressions and the time of occurrences in the time series data. 

Using  the  mean as the normalization threshold, the software  determines the event amplitude with the first subroutine and time of occurrence of the reductions with the second subroutine from the  CR data simultaneously. The output of the present algorithm makes it possible to select FDs as low as -0.05(\%). As noted in \citet{ok:2019}, the number of FDs detected depends on the baseline used. Using a very small baseline ($\mathrm{CR} (\%)\leq -0.01$), our algorithm selected  230  FDs  from Oulu  neutron monitor station covering the period 1998-2002. This result is presented in Table \ref{table 3} with the associated solar wind data and geomagnetic storm indices. The automated program   is written in R, a non-commercial software \citep{R:2014}. 

R is one of the fastest developing statistical computer program. The software is generally referred to as R Foundation for Statistical Computing Platform. The project was initiated by Robert Gentleman and Ross Ihaka of the statistical department of  University of Auckland. Readers are referred to \citet{ok:2021} for details on the current R code.

Out of the 230 algorithm-selected FD datasets, we formed a subset of 129 large FDs (CR(\%) $\leq-3$) catalog. The two FD catalogues with their associated  solar wind data and geomagnetic storm indices are presented in Tables 3  and 4. In the Tables, S/N stands for serial number, Date  of ocurrences of both the FDs and the solar wind/geophysical activity characteristics is in column 2, IMF stand for interplanetary magnetic field intensity, SWS represent solar wind speed,  kp, Dst, ap are geomagnetic storm indices while solar wind  disturbance characteristic SWSxIMF is in the last column.  
\section{Analysis and  results}
The variability of CR flux is highly unpredictable. At the moment, no method of FD selection either semi-automated or automated, can address comprehensively the diversities of FD features. Validation of FD data list is crucial in tackling  factors that could potentially introduce feigned intensity depressions in CR flux. It also help to check  the  efficiency  of the computer software. 

In line with the relationship that theoretically exist between FD, solar wind structure and geomagnetic activity indices as pointed out in section 1 \citep[e.g.][]{be:05, be:08, lin:2016}, we attempt to validate the current  FD data by comparing the amplitudes of the FD catalogue with  the solar wind data and geomagnetic storm indices. This was carried out with the Pearson's Product Moment Correlation statistics as presented in  section 5.2.

The present analysis will focus on the 129 large FD (CR(\%) $\leq-3$) catalog presented in  Table \ref{table 3}. Investigations on the remaining  FD data in Table \ref{table 2}  will be taken up in future work. Large FDs (CR(\%) $\leq-3$) is often adopted as the common threshold to select large FDs  \citep[e.g.][]{kris:08, Ka:10, ok:2020e}. The  matrix of correlation results for  large FDs with corresponding solar wind  and geomagnetic activity index data   is given in Table \ref{table 1}. The columns of the  table are explained thus: S/N stands for serial number, Parameters represent each of the two continuous variables, $R^2$ indicate coefficient of determination, r is the Pearson's product moment correlation coefficient   and p-value is chance probability. 

\begin{figure}[h!]
 \centering
  \includegraphics[width=1.36\textwidth]{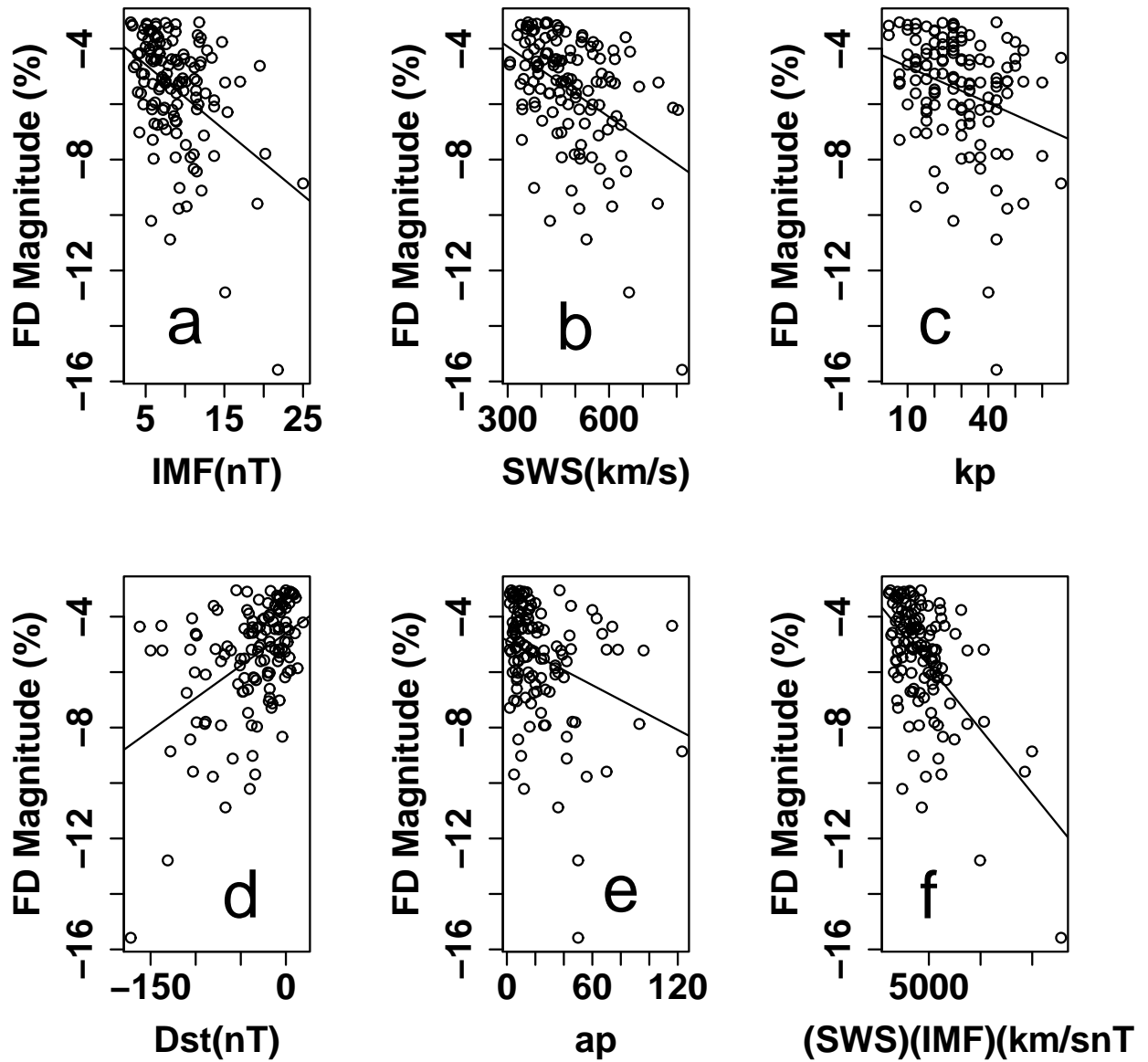}
 \caption{Scatter Plots of Large FDs CR(\%) $\leq-3$) and Related Solar Wind Parameters and Geomagnetic Activity Indices}
\label{Figure1}
 \end{figure}

The scatter plot of FD magnitude and IMF intensity is displayed in panel \textbf{a} of Figure \ref{Figure1}. The results of the regression and correlation analyses for the FD-IMF relation yield $R^2$ $\sim{0.19}$, $r\sim{-0.44}$ with p-value of $1.98\times 10^{-07}$. The correlation is statistically significant at 95\% confidence level. $R^2$, which is the coefficient of determination expresses the percentage of variation in the regressand (dependent variable) that can be attributed \citep{oki:20} to the regressor (independent or explanatory variable). The regression result hence imply  that  19\% of CR intensity variation can be accounted for by IMF intensity in  the large FDs (CR(\%) $\leq-3$) dataset in Table .

The graph of FD magnitude and SWS   is presented in panel \textbf{b} of Figure \ref{Figure1}. The analysis of the FD-SWS relation give  $R^2$, r and p-value as  0.19, -0.44, $1.22\times 10^{-07}$. The p-value indicate that the result is  statistically significant at 95\% confidence level. The regression analysis  suggest that   19\% of GCR density changes in the  sample  in Table \ref{table 3} is driven by high speed solar wind.

Scatter plot of FD amplitude and kp relation for the  dataset of Table \ref{table 3} is plotted in panel \textbf{c} of Figure \ref{Figure1}. The  correlation and regression analyses of FD-kp relation yield $R^2$, r and p-value are respectively 0.08, -0.28, $1.23\times 10^{-05}$. The p-value indicates a high level of significance. $R^2$ value in the  dataset predicts that  8\% of FD is caused by geomagnetic index (kp). 

The plot of FD magnitude and Dst from Table \ref{table 3} is shown in panel \textbf{d} of Figure \ref{Figure1}. The regression analysis for the FD-Dst parameters gives  $R^2$, coefficient of correlation r and p-value  as 0.21, 0.46, $3.51\times 10^{-08}$ respectively. High statistical significance is implied here from the chance probability value. From the observed $R^2$ result, we infer that  21\%  of cosmic ray variation is related to disturbance storm-time index corresponding to  data in Table \ref{table 3}.

We show FD magnitude and ap graph for the  dataset of Table \ref{table 3} in panel \textbf{e} of Figure \ref{Figure1}. The corresponding $R^2$, r and p-value from the FD-ap relation analysis results are  0.08, 0.29, $4.95\times 10^{-04}$ for data in Table \ref{table 3}. While the results here is statistically significant from the chance probability value, the $R^2$ value suggests that  only  8\% of the variations in FD events can be attributed to ap in the large FD data list. 

In panel \textbf{f} of Figure \ref{Figure1}, we present the scatter plot of FD magnitude and ($SWS\times{IMF}$) for the  129 FD dataset. The result of our regression and correlation analyses give $R^2$, r and p-value respectively as  0.34, 0.58, $4.65\times 10^{-13}$. This result is statistically significant at 95\% confidence level. $R^2$ value obtained here suggest that  34\% of cosmic ray variability can be accounted for by solar wind disturbance characteristic. 

\section{Discussion and conclusion} 
\subsection{Discussion}
The results of the statistical  analysis reveal that, on average, large amplitude FDs (CR(\%) $\leq-3$) show  pronounced tendencies  of linear relationship with  solar wind data and geomagnetic storm activity parameters. 

 Large amplitude FDs are generally presumed to be caused by coronal mass ejections. The statistical level of significance of  correlations between FDs, solar wind and geophysical parameters appear to suggest that SWS, IMF intensity and storm-time disturbance 
index (Dst) are key drivers of  variations of FDs, which  in turn are linked to  CMEs \citep{richard:2004, richard:2011, be:2014}. 

Using a population of FDs CR(\%) ${>}$ 4) in magnitude that occurred between 1966 and 1972, \citet{barn:1973} observed that the interplanetary disturbances responsible for Forbush decreases are characterized by a region of high solar wind velocity and enhanced interplanetary magnetic field intensity.
Similarly, \citet{ba:75} reported an association between FDs and variations in IMF intensity using Deep River neutron monitor data  from 1967 to 1968. They noted that from 1967 to 1968, all the cosmic ray flux variation is associated with  the passage of a region when magnetic field intensity is high.  About two decades later, \citet{if:1998} found that both sudden increase in IMF and the SWS are preceded by the onset of the Forbush decrease at Deeep River neutron monitor (NM) station. The results of the present work in which FD-IMF and  FD-SWS relations yield correlations significant at 95\% confidence level is in close agreement with results in the literature referred to above. From statistical analysis of 695 events selected with GSM \citet{belov:2001} found that 49\% of FD variation are due to the effect of solar disturbance characteristic.
The result obtained here for FD-solar disturbance characteristic is in fair agreement with their finding. 

 Recently, \citet{lin:2016} showed that FD amplitude do not correlate with Dst  using 68 large (CR(\%) ${>}$ 2) GSM-selected FDs. They reported evidence of correlation for the FD-SWS relation from the same sample with coefficient of determination $R^2$ = 0.16. This result could mean  that 16\% of the modulation in the CR intensity is related to the effect of solar wind structure. It is interesting to note that the current statistical analysis indicates higher values of coefficient of determination. The improved relation could be a pointer to the differences in the semi-automated and the current fully automated FD event identification approaches.

\citet{bad:2015} found FD-SWS relation of r $\sim {0.58}$. The FD-SWS correlation r $\sim {0.44}$ obtained in the present investigation is in    agreement with their result. Our results for FD-SWS (r $\sim {0.44}$) and FD-IMF (r $\sim {0.44}$) does not reflect the submission of  \citet{ok:2020e} who obtained r $\sim {0.81}$ and  $\sim{0.34}$ respectively. This could be as a result of the data periods. Our data period is from 1998 to 2002, while the period of the FD selection by \citet{ok:2020e} is 2003 to 2005 which is included the period of high solar activity.  However, from the probability values, the results obtained here is statistically significant, hence, we infer that  SWS and IMF might play  key roles in CR intensty modulation. 

Using a large FD event database of 5900 that span from 1957 to 2006, \citet{be:08} submitted that large amplitude  FDs (CR(\%) $<-5$) are associated with strong geomagnetic storms. The  results obtained here for the association between FD magnitude and geomagnetic storm indices  significant at 95\% confidence level are in agreement with their submission.
\subsection{Conclusion}
From the  study of 129 algorithm selected large FDs (CR(\%) $\leq-3$) for cosmic ray intensity at Oulu NM station between 1998 and 2002, the following conclusions
can be drawn:
\begin{itemize}
\item  A new automated FD location code has been succesfully developed and employed to select FD list from the daily averaged  CR raw data  at Oulu  NM station from 1998 to 2002.  
\item The computer program-selected FD was validated by comparing its amplitude with the corresponding solar wind data and geomagnetic storm activity indices.
 \item The   analysis shows that the correlations between the continous variables are statistically significant and hence reliable. This results reveal that the selected FDs are not spurious. 
\item We infer that IMF intensity, SWS, geomagnetic storm indices (Dst, kp and ap) are key drivers in galactic cosmic ray modulation.
\end{itemize}

   
\begin{table}[ht]
\caption{Correlation Results with Large  FDs CR(\%) $\leq-3$)}
\label{table 1}
        \centering
        \begin{tabular}{rlrlrl}
            \hline
            S/N & Parameters & $R^2$ & r & p-values \\
            \hline
             \hline
1 & FD-IMF & 0.19 &  -0.44 & $1.98\times 10^{-07}$ \\
2 & FD-SWS & 0.19 &  -0.44 & $1.22\times 10^{-07}$ \\ 
  3 & FD-kp &  0.08 &  -0.28 & $1.23\times 10^{-05}$ \\ 
  4 & FD-Dst & 0.21 &  0.46 & $3.51\times 10^{-08}$ \\  
  5 & FD-ap & 0.08 &  0.29 & $4.95\times 10^{-04}$ \\ 
  6 & FD-SWSxIMF & 0.34 &  0.58 & $4.65\times 10^{-13}$ \\ 
            \hline
        \end{tabular}
\end{table}

\begin{table}[ht]
\caption{ \textbf{Oulu NM station FDs (All FDs) and Associated  Solar Wind Data and Geomagnetic Activity Indices  from 1998-2002}}
\label{table 2}
\centering
\begin{tabular}{rlrrrrrrr}
  \hline
 S/N & Date & FD(\%) & IMF (nT) & SWS(km$s^{-1}$) & kp & Dst (nT) & ap & SWSx IMF(km$s^{-1}$)(nT) \\ 
  \hline
1 & 1998-08-27 & -2.62 & 14.10 & 630 &  70 & -129 & 144 & 8883.00 \\ 
  2 & 1998-09-25 & -0.75 & 18.00 & 713 &  60 & -118 & 117 & 12834.00 \\ 
  3 & 1999-01-24 & -0.67 & 7.60 & 517 &  30 & -38 &  17 & 3929.20 \\ 
  4 & 1999-02-18 & -2.23 & 17.10 & 599 &  60 & -84 &  80 & 10242.90 \\ 
  5 & 1999-08-22 & -0.53 & 6.00 & 428 &  20 & -27 &  12 & 2568.00 \\ 
  6 & 1999-10-17 & -0.90 & 5.20 & 520 &  33 & -33 &  21 & 2704.00 \\ 
  7 & 1999-10-22 & -0.72 & 14.20 & 608 &  57 & -134 &  91 & 8633.60 \\ 
  8 & 1999-10-24 & -0.85 & 5.80 & 571 &  40 & -58 &  26 & 3311.80 \\ 
  9 & 1999-11-12 & -0.18 & 4.90 & 554 &  23 & -34 &  10 & 2714.60 \\ 
  10 & 1999-11-17 & -0.64 & 10.60 & 447 &  27 & -30 &  12 & 4738.20 \\ 
  11 & 1999-11-20 & -1.06 & 8.10 & 443 &  20 & -16 &   9 & 3588.30 \\ 
  12 & 1999-11-22 & -0.87 & 9.80 & 453 &  23 & -12 &  11 & 4439.40 \\ 
  13 & 1999-12-03 & -1.53 & 12.70 & 425 &  30 &  -7 &  16 & 5397.50 \\ 
  14 & 1999-12-13 & -5.51 & 11.40 & 489 &  33 & -46 &  26 & 5574.60 \\ 
  15 & 1999-12-27 & -2.70 & 7.90 & 410 &  17 &   2 &   7 & 3239.00 \\ 
  16 & 1999-12-31 & -0.75 & 10.80 & 653 &  43 & -18 &  36 & 7052.40 \\ 
  17 & 2000-01-02 & -0.70 & 5.10 & 679 &  30 & -19 &  16 & 3462.90 \\ 
  18 & 2000-01-04 & -0.79 & 5.80 & 577 &  27 & -15 &  13 & 3346.60 \\ 
  19 & 2000-01-06 & -0.75 & 6.90 & 533 &  33 & -22 &  19 & 3677.70 \\ 
  20 & 2000-01-24 & -0.82 & 10.10 & 366 &  27 & -40 &  13 & 3696.60 \\ 
  21 & 2000-01-29 & -0.05 & 5.60 & 722 &  43 & -26 &  30 & 4043.20 \\ 
  22 & 2000-02-07 & -0.55 & 5.40 & 629 &  43 & -35 &  31 & 3396.60 \\ 
  23 & 2000-02-12 & -3.76 & 14.70 & 553 &  50 & -76 &  60 & 8129.10 \\ 
  24 & 2000-02-21 & -2.13 & 14.30 & 423 &  33 &  -1 &  21 & 6048.90 \\ 
  25 & 2000-03-01 & -2.73 & 7.60 & 480 &  33 & -22 &  21 & 3648.00 \\ 
  26 & 2000-03-04 & -1.83 & 5.20 & 361 &   7 &   0 &   3 & 1877.20 \\ 
  27 & 2000-03-06 & -1.59 & 8.90 & 411 &  27 &  -8 &  12 & 3657.90 \\ 
  28 & 2000-03-09 & -1.81 & 6.50 & 391 &  10 & -14 &   4 & 2541.50 \\ 
29 & 2000-03-13 & -2.62 & 3.50 & 366 &  10 & -13 &   4 & 1281.00 \\
  30 & 2000-03-19 & -2.26 & 8.30 & 366 &  20 &   2 &   7 & 3037.80 \\
  31 & 2000-03-24 & -3.59 & 6.60 & 649 &  23 &  -3 &  11 & 4283.40 \\
  32 & 2000-03-30 & -3.15 & 5.20 & 446 &  27 &  -2 &  12 & 2319.20 \\ 
  33 & 2000-04-05 & -2.99 & 7.30 & 387 &  20 & -34 &  11 & 2825.10 \\ 
  34 & 2000-04-07 & -4.36 & 9.90 & 573 &  50 & -162 &  74 & 5672.70 \\ 
  35 & 2000-04-14 & -0.79 & 6.60 & 318 &   3 &  -2 &   2 & 2098.80 \\ 
  36 & 2000-04-17 & -1.76 & 6.20 & 457 &  23 & -23 &  12 & 2833.40 \\ 
  37 & 2000-04-19 & -1.98 & 10.10 & 461 &  23 &  -7 &  12 & 4656.10 \\ 
  38 & 2000-04-22 & -1.79 & 5.50 & 449 &  13 &   1 &   5 & 2469.50 \\ 
  39 & 2000-04-24 & -2.30 & 9.40 & 485 &  33 & -25 &  21 & 4559.00 \\ 
  40 & 2000-05-03 & -3.59 & 6.20 & 520 &  30 & -12 &  17 & 3224.00 \\ 
  41 & 2000-05-08 & -4.21 & 9.80 & 360 &  10 &  19 &   4 & 3528.00 \\ 
  42 & 2000-05-16 & -3.71 & 8.40 & 444 &  30 &  -8 &  16 & 3729.60 \\ 
  43 & 2000-05-24 & -7.87 & 13.70 & 636 &  60 & -90 &  93 & 8713.20 \\ 
  44 & 2000-05-30 & -4.38 & 6.20 & 617 &  37 & -29 &  22 & 3825.40 \\ 
  45 & 2000-06-09 & -9.69 & 10.20 & 609 &  13 & -34 &   5 & 6211.80 \\ 
  46 & 2000-06-20 & -5.94 & 6.20 & 379 &  17 &   5 &   6 & 2349.80 \\ 
 \hline
\end{tabular}
\end{table}

 \begin{table}[ht]
\label*{table 2}
\centering
\begin{tabular}{rlrrrrrrr}
  \hline
 S/N & Date & FD(\%) & IMF (nT) & SWS (km$s^{-1}$) & kp & Dst (nT) & ap & SWSx IMF(km$s^{-1}$)(nT) \\ 
  \hline
  47 & 2000-06-24 & -6.01 & 9.00 & 551 &  27 & -20 &  15 & 4959.00 \\ 
  48 & 2000-06-26 & -6.19 & 11.50 & 512 &  43 & -36 &  40 & 5888.00 \\ 
  49 & 2000-07-02 & -3.32 & 5.50 & 378 &  10 &  11 &   4 & 2079.00 \\ 
  50 & 2000-07-05 & -3.34 & 5.80 & 449 &  20 &   1 &   9 & 2604.20 \\ 
  51 & 2000-07-08 & -3.20 & 6.30 & 363 &  17 &   9 &   6 & 2286.90 \\ 
  52 & 2000-07-11 & -5.86 & 13.70 & 458 &  43 &  13 &  34 & 6274.60 \\ 
  53 & 2000-07-13 & -8.33 & 11.10 & 573 &  37 &  -4 &  42 & 6360.30 \\ 
  54 & 2000-07-16 & -15.58 & 21.80 & 816 &  43 & -172 &  50 & 17788.80 \\ 
  55 & 2000-07-20 & -10.88 & 8.10 & 533 &  43 & -67 &  36 & 4317.30 \\ 
  56 & 2000-07-22 & -10.21 & 5.70 & 425 &  27 & -40 &  12 & 2422.50 \\ 
  57 & 2000-07-29 & -7.92 & 8.80 & 460 &  37 & -38 &  27 & 4048.00 \\ 
  58 & 2000-08-06 & -7.97 & 6.00 & 515 &  30 & -32 &  16 & 3090.00 \\ 
  59 & 2000-08-12 & -8.86 & 25.00 & 599 &  67 & -128 & 123 & 14975.00 \\
  60 & 2000-08-26 & -4.46 & 6.00 & 385 &  13 &  -8 &   5 & 2310.00 \\ 
  61 & 2000-08-31 & -3.93 & 5.00 & 546 &  27 & -18 &  14 & 2730.00 \\ 
  62 & 2000-09-03 & -4.18 & 6.80 & 413 &  17 & -14 &   7 & 2808.40 \\ 
  63 & 2000-09-09 & -5.61 & 4.40 & 425 &  13 &  -9 &   5 & 1870.00 \\ 
  64 & 2000-09-15 & -5.47 & 6.70 & 363 &  20 &   4 &  12 & 2432.10 \\ 
  65 & 2000-09-18 & -9.59 & 19.20 & 744 &  53 & -103 &  70 & 14284.80 \\ 
  66 & 2000-09-26 & -3.89 & 5.90 & 566 &  37 & -41 &  24 & 3339.40 \\ 
  67 & 2000-09-29 & -4.01 & 5.50 & 378 &  17 & -19 &   7 & 2079.00 \\ 
  68 & 2000-10-01 & -4.16 & 4.30 & 418 &  27 & -37 &  13 & 1797.40 \\ 
  69 & 2000-10-05 & -4.33 & 13.40 & 486 &  67 & -138 & 116 & 6512.40 \\ 
  70 & 2000-10-07 & -4.58 & 3.80 & 391 &  10 & -36 &   4 & 1485.80 \\ 
  71 & 2000-10-14 & -3.61 & 12.00 & 411 &  47 & -80 &  45 & 4932.00 \\ 
  72 & 2000-10-20 & -1.81 & 4.90 & 433 &   7 &  -2 &   3 & 2121.70 \\ 
  73 & 2000-10-22 & -1.71 & 8.70 & 463 &  27 &  -6 &  16 & 4028.10 \\ 
  74 & 2000-10-29 & -6.09 & 13.70 & 381 &  40 & -89 &  34 & 5219.70 \\ 
  75 & 2000-11-04 & -4.48 & 12.00 & 436 &  40 & -22 &  26 & 5232.00 \\ 
  76 & 2000-11-07 & -7.79 & 20.20 & 512 &  43 & -89 &  46 & 10342.40 \\ 
  77 & 2000-11-11 & -6.21 & 7.20 & 804 &  30 & -35 &  16 & 5788.80 \\ 
  78 & 2000-11-13 & -4.90 & 4.50 & 578 &  23 & -18 &   9 & 2601.00 \\ 
  79 & 2000-11-16 & -4.92 & 4.80 & 391 &   7 &   2 &   3 & 1876.80 \\ 
  80 & 2000-11-24 & -4.40 & 8.80 & 406 &  23 &   0 &  11 & 3572.80 \\ 
  81 & 2000-11-29 & -9.77 & 9.20 & 512 &  47 & -81 &  56 & 4710.40 \\ 
  82 & 2000-12-06 & -6.19 & 5.80 & 343 &  17 &  -1 &   7 & 1989.40 \\ 
  83 & 2000-12-11 & -4.20 & 4.00 & 552 &  20 &   0 &   8 & 2208.00 \\ 
  84 & 2000-12-15 & -3.17 & 3.30 & 356 &   3 &   8 &   2 & 1174.80 \\ 
  85 & 2000-12-19 & -3.29 & 4.80 & 356 &  13 &  -7 &   5 & 1708.80 \\ 
  86 & 2000-12-23 & -4.58 & 9.20 & 306 &  33 & -38 &  21 & 2815.20 \\ 
  87 & 2000-12-25 & -4.60 & 11.90 & 352 &  17 &  10 &   6 & 4188.80 \\ 
 88 & 2000-12-27 & -4.90 & 7.70 & 390 &  20 &  -1 &   8 & 3003.00 \\
  89 & 2000-12-31 & -3.51 & 4.60 & 329 &   3 &   4 &   2 & 1513.40 \\ 
  90 & 2001-01-03 & -3.78 & 6.80 & 351 &  20 &  -8 &   8 & 2386.80 \\ 
  91 & 2001-01-10 & -3.69 & 5.30 & 367 &  13 &   0 &   5 & 1945.10 \\ 
  92 & 2001-01-14 & -3.37 & 5.40 & 372 &  20 &  -8 &   8 & 2008.80 \\ 
  93 & 2001-01-19 & -3.14 & 3.30 & 339 &  10 &   7 &   4 & 1118.70 \\ 
  94 & 2001-01-25 & -4.67 & 3.50 & 387 &  13 & -27 &   6 & 1354.50 \\ 
  95 & -2001-01-31 & 2.92 & 7.60 & 409 &  30 & -22 &  18 & 3108.40 \\
\hline
\end{tabular}
\end{table}

\begin{table}[ht]
\label*{table 2}
\centering
\begin{tabular}{rlrrrrrrr}
  \hline
 S/N & Date & FD(\%) & IMF (nT) & SWS (km$s^{-1}$) & kp & Dst (nT) & ap & SWSxIMF(km$s^{-1}$)(nT) \\ 
  \hline 
  96 & 2001-02-07 & -1.84 & 6.00 & 488 &  17 &  -6 &   6 & 2928.00 \\ 
  97 & 2001-02-11 & -1.24 & 5.30 & 398 &  17 &   0 &   6 & 2109.40 \\ 
  98 & 2001-02-14 & -2.03 & 5.80 & 513 &  33 & -34 &  19 & 2975.40 \\ 
  99 & 2001-02-20 & -1.22 & 6.90 & 315 &  17 &   5 &   7 & 2173.50 \\ 
  100 & 2001-03-04 & -1.46 & 6.90 & 448 &  33 & -17 &  19 & 3091.20 \\ 
  101 & 2001-03-21 & -0.57 & 12.10 & 320 &  17 & -65 &   8 & 3872.00 \\ 
  102 & 2001-03-28 & -2.73 & 8.90 & 608 &  43 & -53 &  44 & 5411.20 \\ 
  103 & 2001-04-01 & -5.22 & 7.50 & 746 &  43 & -137 &  38 & 5595.00 \\ 
  104 & 2001-04-05 & -5.25 & 7.50 & 617 &  33 & -31 &  19 & 4627.50 \\ 
  105 & 2001-04-09 & -6.43 & 8.60 & 622 &  33 & -53 &  20 & 5349.20 \\ 
  106 & 2001-04-12 & -12.79 & 15.10 & 659 &  40 & -131 &  50 & 9950.90 \\ 
  107 & 2001-04-16 & -5.57 & 4.00 & 453 &  20 & -24 &   8 & 1812.00 \\ 
  108 & 2001-04-19 & -4.60 & 7.90 & 436 &  17 & -41 &   6 & 3444.40 \\ 
  109 & 2001-04-22 & -3.05 & 11.80 & 360 &  43 & -55 &  37 & 4248.00 \\ 
  110 & 2001-04-26 & -1.95 & 6.80 & 438 &  17 &  -6 &   6 & 2978.40 \\ 
  111 & 2001-04-29 & -6.92 & 7.60 & 596 &  23 & -18 &  13 & 4529.60 \\ 
  112 & 2001-05-08 & -1.63 & 9.20 & 410 &  27 & -20 &  16 & 3772.00 \\ 
  113 & 2001-05-12 & -1.93 & 10.80 & 534 &  40 & -35 &  30 & 5767.20 \\ 
  114 & 2001-05-16 & -1.42 & 5.10 & 487 &  17 &  -9 &   7 & 2483.70 \\ 
  115 & 2001-05-20 & -0.84 & 5.40 & 384 &  13 &   0 &   5 & 2073.60 \\ 
  116 & 2001-05-25 & -2.95 & 6.60 & 557 &  17 &   5 &   7 & 3676.20 \\ 
  117 & 2001-05-28 & -5.30 & 9.10 & 505 &  33 &  -8 &  18 & 4595.50 \\   
 118 & 2001-06-03 & -2.11 & 5.30 & 508 &  17 &  -7 &   6 & 2692.40 \\
  119 & 2001-06-09 & -1.88 & 9.70 & 510 &  33 &   0 &  25 & 4947.00 \\
  120 & 2001-06-12 & -1.83 & 5.20 & 433 &  10 &  -2 &   4 & 2251.60 \\ 
  121 & 2001-06-20 & -1.93 & 5.90 & 700 &  23 & -20 &  11 & 4130.00 \\ 
  122 & 2001-06-26 & -1.26 & 6.20 & 464 &  27 &  -2 &  13 & 2876.80 \\ 
  123 & 2001-06-29 & -1.17 & 3.40 & 347 &   7 &  11 &   3 & 1179.80 \\ 
  124 & 2001-07-05 & -2.16 & 7.00 & 423 &  27 &   5 &  12 & 2961.00 \\ 
  125 & 2001-07-09 & -1.58 & 4.10 & 427 &  17 &  -8 &  10 & 1750.70 \\ 
  126 & 2001-07-17 & -0.27 & 8.90 & 592 &  30 & -13 &  16 & 5268.80 \\ 
  127 & 2001-07-19 & -0.53 & 4.80 & 605 &  17 &   0 &   7 & 2904.00 \\ 
  128 & 2001-07-21 & -0.32 & 5.60 & 412 &  10 &  11 &   4 & 2307.20 \\ 
  129 & 2001-07-26 & -1.63 & 4.60 & 552 &  23 & -15 &  10 & 2539.20 \\ 
  130 & 2001-07-30 & -1.51 & 6.80 & 312 &  17 &  10 &   7 & 2121.60 \\ 
  131 & 2001-08-03 & -2.43 & 7.20 & 405 &  27 &   0 &  12 & 2916.00 \\ 
  132 & 2001-08-06 & -2.50 & 7.00 & 440 &  30 & -18 &  17 & 3080.00 \\ 
  133 & 2001-08-14 & -1.56 & 7.80 & 456 &  23 & -10 &  10 & 3556.80 \\ 
  134 & 2001-08-18 & -3.76 & 11.80 & 518 &  27 & -43 &  15 & 6112.40 \\ 
  135 & 2001-08-24 & -3.05 & 3.10 & 414 &   7 &   0 &   3 & 1283.40 \\ 
  136 & 2001-08-29 & -7.02 & 4.20 & 459 &  13 &  -7 &   5 & 1927.80 \\ 
  137 & 2001-09-07 & -2.72 & 6.10 & 369 &   7 &   6 &   3 & 2250.90 \\ 
  138 & 2001-09-15 & -1.69 & 7.60 & 541 &  33 & -11 &  20 & 4111.60 \\ 
  139 & 2001-09-19 & -1.46 & 6.50 & 422 &  20 &  -5 &   9 & 2743.00 \\ 
  140 & 2001-09-22 & -0.92 & 5.70 & 327 &  20 &   1 &   8 & 1863.90 \\ 
  141 & 2001-09-26 & -7.92 & 10.70 & 549 &  33 & -72 &  26 & 5874.30 \\ 
  142 & 2001-10-01 & -7.81 & 11.10 & 498 &  47 & -99 &  48 & 5527.80 \\ 
  143 & 2001-10-09 & -4.38 & 8.30 & 445 &  30 & -37 &  18 & 3693.50 \\ 
  144 & 2001-10-12 & -5.76 & 11.40 & 501 &  40 & -51 &  34 & 5711.40 \\
\hline
\end{tabular}
\end{table}

\begin{table}[ht]
\label*{table 2}
\centering
\begin{tabular}{rlrrrrrrr}
  \hline
 S/N & Date & FD(\%) & IMF (nT) & SWS (km$s^{-1}$) & kp & Dst (nT) & ap & SWSx IMF(km$s^{-1}$)(nT) \\ 
  \hline
 
  145 & 2001-10-22 & -5.22 & 15.10 & 578 &  60 & -150 &  96 & 8727.80 \\ 
  146 & 2001-10-28 & -4.68 & 11.20 & 450 &  47 & -99 &  44 & 5040.00 \\ 
  147 & 2001-11-07 & -6.75 & 6.50 & 635 &  30 & -110 &  19 & 4127.50 \\ 
148 & 2001-11-15 & -2.16 & 6.70 & 300 &  17 &  -5 &   9 & 2010.00 \\
  149 & 2001-11-22 & -3.07 & 7.50 & 418 &  23 & -17 &   9 & 3135.00 \\ 
  150 & 2001-11-25 & -8.43 & 11.50 & 650 &  20 & -106 &   8 & 7475.00 \\ 
  151 & 2001-12-01 & -1.46 & 8.00 & 348 &  17 &   0 &   7 & 2784.00 \\ 
  152 & 2001-12-06 & -3.51 & 6.40 & 432 &  27 & -20 &  11 & 2764.80 \\ 
  153 & 2001-12-17 & -3.39 & 8.80 & 471 &  30 & -30 &  16 & 4144.80 \\ 
  154 & 2001-12-22 & -1.02 & 7.70 & 379 &  20 & -40 &   8 & 2918.30 \\ 
  155 & 2001-12-24 & -0.67 & 9.00 & 508 &  37 & -30 &  23 & 4572.00 \\ 
  156 & 2001-12-27 & -1.29 & 7.30 & 369 &  17 & -16 &   7 & 2693.70 \\ 
  157 & 2001-12-29 & -2.25 & 15.40 & 397 &  27 &  34 &  11 & 6113.80 \\ 
  158 & 2002-01-03 & -7.29 & 5.90 & 342 &   7 & -16 &   2 & 2017.80 \\ 
  159 & 2002-01-11 & -6.63 & 8.90 & 610 &  40 & -42 &  27 & 5429.00 \\ 
  160 & 2002-01-19 & -3.12 & 8.80 & 372 &  27 &   4 &  14 & 3273.60 \\ 
  161 & 2002-01-21 & -3.24 & 7.90 & 452 &  27 & -10 &  11 & 3570.80 \\ 
  162 & 2002-01-24 & -2.99 & 5.90 & 360 &  10 & -12 &   4 & 2124.00 \\ 
  163 & 2002-01-30 & -5.27 & 5.90 & 338 &   7 &  -1 &   3 & 1994.20 \\ 
  164 & 2002-02-01 & -5.09 & 11.20 & 347 &  27 & -17 &  14 & 3886.40 \\ 
  165 & 2002-02-11 & -0.77 & 8.10 & 491 &  27 & -17 &  14 & 3977.10 \\ 
  166 & 2002-02-15 & -0.42 & 6.30 & 371 &   7 &  -3 &   3 & 2337.30 \\ 
  167 & 2002-02-19 & -0.75 & 7.70 & 401 &  13 & -17 &   6 & 3087.70 \\ 
  168 & 2002-02-23 & -1.89 & 6.40 & 362 &  10 &  -6 &   4 & 2316.80 \\ 
  169 & 2002-03-02 & -2.26 & 9.90 & 386 &  13 & -14 &   6 & 3821.40 \\ 
  170 & 2002-03-05 & -1.83 & 9.50 & 646 &  33 & -22 &  21 & 6137.00 \\ 
  171 & 2002-03-12 & -1.76 & 7.90 & 453 &  27 &  -3 &  11 & 3578.70 \\ 
  172 & 2002-03-16 & -1.63 & 6.30 & 310 &   7 &  13 &   3 & 1953.00 \\ 
  173 & 2002-03-22 & -5.14 & 7.50 & 444 &  13 &  -4 &   6 & 3330.00 \\ 
  174 & 2002-03-25 & -6.29 & 15.40 & 433 &  17 & -33 &   7 & 6668.20 \\ 
  175 & 2002-03-30 & -3.51 & 11.60 & 521 &  33 &  -5 &  20 & 6043.60 \\ 
  176 & 2002-04-05 & -1.74 & 6.50 & 402 &   7 &   6 &   3 & 2613.00 \\ 
  177 & 2002-04-12 & -2.72 & 8.70 & 432 &  30 &  -1 &  16 & 3758.40 \\ 
178 & 2002-04-15 & -1.58 & 8.80 & 357 &  13 &  -8 &   6 & 3141.60 \\
  179 & 2002-04-18 & -4.06 & 12.80 & 485 &  53 & -104 &  63 & 6208.00 \\ 
  180 & 2002-04-20 & -5.19 & 10.10 & 563 &  53 & -106 &  70 & 5686.30 \\ 
  181 & 2002-04-24 & -5.10 & 7.10 & 488 &  17 & -30 &   7 & 3464.80 \\ 
  182 & 2002-05-04 & -0.69 & 5.10 & 378 &  13 &   4 &   4 & 1927.80 \\ 
  183 & 2002-05-08 & -1.11 & 8.50 & 366 &  20 & -19 &   8 & 3111.00 \\ 
  184 & 2002-05-13 & -2.80 & 6.00 & 457 &  17 & -21 &   8 & 2742.00 \\ 
  185 & 2002-05-15 & -3.10 & 6.30 & 411 &  27 & -43 &  12 & 2589.30 \\ 
  186 & 2002-05-20 & -4.45 & 9.00 & 453 &  23 & -12 &  10 & 4077.00 \\ 
  187 & 2002-05-23 & -5.19 & 17.00 & 606 &  47 & -38 &  78 & 10302.00 \\ 
  188 & 2002-05-28 & -4.11 & 5.60 & 662 &  23 & -26 &   9 & 3707.20 \\ 
  189 & 2002-06-04 & -2.20 & 5.90 & 441 &  27 & -23 &  13 & 2601.90 \\ 
 
 \hline
\end{tabular}
\end{table}

\begin{table}[ht]
\label*{table 2}
\centering
\begin{tabular}{rlrrrrrrr}
  \hline
 S/N & Date & FD(\%) & IMF (nT) & SWS (km$s^{-1}$) & kp & Dst (nT) & ap & SWSx IMF(km$s^{-1}$)(nT) \\ 
  \hline
  190 & 2002-06-07 & -2.26 & 5.00 & 308 &  13 &  -3 &   5 & 1540.00 \\ 
  191 & 2002-06-12 & -2.84 & 6.80 & 368 &  17 & -14 &   6 & 2502.40 \\ 
  192 & 2002-06-16 & -1.86 & 8.80 & 382 &  20 &   6 &   7 & 3361.60 \\ 
  193 & 2002-06-19 & -2.94 & 10.60 & 468 &  27 &  -2 &  11 & 4960.80 \\ 
  194 & 2002-06-24 & -1.32 & 7.10 & 474 &  13 &  -3 &   5 & 3365.40 \\ 
  195 & 2002-06-29 & -0.58 & 5.10 & 344 &  13 &   6 &   5 & 1754.40 \\ 
  196 & 2002-07-03 & -1.79 & 3.80 & 361 &  10 &   1 &   4 & 1371.80 \\ 
  197 & 2002-07-11 & -3.10 & 5.10 & 386 &  13 &   5 &   5 & 1968.60 \\ 
  198 & 2002-07-18 & -4.38 & 5.70 & 441 &  13 &  -8 &   5 & 2513.70 \\ 
  199 & 2002-07-20 & -6.13 & 7.40 & 789 &  30 & -20 &  18 & 5838.60 \\ 
  200 & 2002-07-23 & -5.20 & 4.80 & 472 &  30 & -12 &  17 & 2265.60 \\ 
  201 & 2002-08-02 & -9.12 & 12.10 & 489 &  43 & -59 &  42 & 5916.90 \\ 
  202 & 2002-08-07 & -6.01 & 4.70 & 343 &  10 &   2 &   4 & 1612.10 \\ 
  203 & 2002-08-09 & -5.45 & 8.20 & 397 &  27 &  -7 &  14 & 3255.40 \\ 
  204 & 2002-08-20 & -6.71 & 7.20 & 479 &  33 & -48 &  30 & 3448.80 \\ 
  205 & 2002-08-23 & -6.60 & 8.80 & 402 &  17 & -18 &   7 & 3537.60 \\ 
  206 & 2002-08-28 & -7.65 & 8.90 & 447 &  17 & -19 &   7 & 3978.30 \\ 
  207 & 2002-09-04 & -5.61 & 12.60 & 422 &  47 & -72 &  42 & 5317.20 \\   
208 & 2002-09-08 & -6.01 & 11.70 & 479 &  33 & -101 &  36 & 5604.30 \\
  209 & 2002-09-11 & -5.22 & 9.60 & 458 &  37 & -61 &  26 & 4396.80 \\ 
  210 & 2002-09-19 & -4.06 & 6.60 & 613 &  27 & -24 &  15 & 4045.80 \\ 
  211 & 2002-09-24 & -5.19 & 9.30 & 376 &   7 &  -8 &   2 & 3496.80 \\ 
  212 & 2002-09-28 & -4.48 & 10.70 & 307 &  13 &   3 &   5 & 3284.90 \\ 
  213 & 2002-10-01 & -4.62 & 19.50 & 388 &  50 & -100 &  67 & 7566.00 \\ 
  214 & 2002-10-03 & -5.17 & 11.50 & 464 &  47 & -78 &  45 & 5336.00 \\ 
  215 & 2002-10-09 & -2.35 & 6.30 & 410 &  33 & -60 &  20 & 2583.00 \\ 
  216 & 2002-10-13 & -2.13 & 6.50 & 301 &  13 & -30 &   5 & 1956.50 \\ 
  217 & 2002-10-21 & -6.03 & 5.80 & 571 &  20 & -18 &   8 & 3311.80 \\ 
  218 & 2002-10-25 & -5.37 & 6.80 & 689 &  43 & -68 &  39 & 4685.20 \\ 
  219 & 2002-11-03 & -5.07 & 9.70 & 478 &  43 & -65 &  35 & 4636.60 \\ 
  220 & 2002-11-05 & -6.21 & 8.40 & 545 &  37 & -46 &  24 & 4578.00 \\ 
  221 & 2002-11-12 & -7.13 & 12.40 & 569 &  30 & -15 &  17 & 7055.60 \\ 
  222 & 2002-11-18 & -9.02 & 9.30 & 378 &  23 & -37 &  10 & 3515.40 \\ 
  223 & 2002-11-25 & -4.43 & 7.00 & 460 &  30 & -46 &  15 & 3220.00 \\ 
  224 & 2002-1127 & -5.52 & 9.80 & 538 &  33 & -50 &  24 & 5272.40 \\ 
  225 & 2002-12-01 & -4.41 & 6.40 & 506 &  33 & -32 &  18 & 3238.40 \\ 
  226 & 2002-12-06 & -3.84 & 7.60 & 394 &  23 &  -6 &   9 & 2994.40 \\ 
  227 & 2002-12-08 & -5.02 & 7.10 & 599 &  27 & -28 &  12 & 4252.90 \\ 
  228 & 2002-12-15 & -5.61 & 8.90 & 502 &  20 & -28 &   8 & 4467.80 \\ 
  229 & 2002-12-20 & -6.70 & 6.10 & 528 &  33 & -47 &  21 & 3220.80 \\ 
  230 & 2002-12-23 & -7.47 & 10.10 & 517 &  37 & -42 &  24 & 5221.70 \\ 
   \hline
\end{tabular}
\end{table}

\begin{table}[ht]
\caption{ \textbf{Oulu NMs Large  FDs (CR(\%)$\leq-3$)) and Associated  Solar Wind parameters and Geomagnetic Activity indices  from 1998-2002}}
\label{table 3}
\centering
\begin{tabular}{rlrrrrrrr}
  \hline
 S/N & Date & FD(\%) & IMF (nT) & SWS (km$s^{-1}$) & kp & Dst (nT) & ap & SWSx IMF(km$s^{-1}$)(nT) \\ 
  \hline

1 & 2001-04-22 & -3.05 & 3.10 & 414 &   7 &   0 &   3 & 1283.40 \\ 
  2 & 2001-08-24 & -3.05 & 11.80 & 360 &  43 & -55 &  37 & 4248.00 \\ 
  3 & 2001-11-22 & -3.07 & 7.50 & 418 &  23 & -17 &   9 & 3135.00 \\ 
  4 & 2002-05-15 & -3.10 & 6.30 & 411 &  27 & -43 &  12 & 2589.30 \\ 
  5 & 2002-07-11 & -3.10 & 5.10 & 386 &  13 &   5 &   5 & 1968.60 \\ 
  6 & 2002-01-19 & -3.12 & 8.80 & 372 &  27 &   4 &  14 & 3273.60 \\ 
  7 & 2001-01-19 & -3.14 & 3.30 & 339 &  10 &   7 &   4 & 1118.70 \\ 
  8 & 2000-03-30 & -3.15 & 5.20 & 446 &  27 &  -2 &  12 & 2319.20 \\ 
  9 & 2000-12-15 & -3.17 & 3.30 & 356 &   3 &   8 &   2 & 1174.80 \\ 
  10 & 2000-07-08 & -3.20 & 6.30 & 363 &  17 &   9 &   6 & 2286.90 \\ 
  11 & 2002-01-21 & -3.24 & 7.90 & 452 &  27 & -10 &  11 & 3570.80 \\ 
  12 & 2000-12-19 & -3.29 & 4.80 & 356 &  13 &  -7 &   5 & 1708.80 \\ 
  13 & 2000-07-02 & -3.32 & 5.50 & 378 &  10 &  11 &   4 & 2079.00 \\ 
  14 & 2000-07-05 & -3.34 & 5.80 & 449 &  20 &   1 &   9 & 2604.20 \\ 
  15 & 2001-01-14 & -3.37 & 5.40 & 372 &  20 &  -8 &   8 & 2008.80 \\ 
  16 & 2001-12-17 & -3.39 & 8.80 & 471 &  30 & -30 &  16 & 4144.80 \\ 
  17 & 2000-12-31 & -3.51 & 6.40 & 432 &  27 & -20 &  11 & 2764.80 \\ 
  18 & 2001-12-06 & -3.51 & 4.60 & 329 &   3 &   4 &   2 & 1513.40 \\ 
  19 & 2002-03-30 & -3.51 & 11.60 & 521 &  33 &  -5 &  20 & 6043.60 \\ 
  20 & 2000-03-24 & -3.59 & 6.60 & 649 &  23 &  -3 &  11 & 4283.40 \\ 
  21 & 2000-05-03 & -3.59 & 6.20 & 520 &  30 & -12 &  17 & 3224.00 \\ 
  22 & 2000-10-14 & -3.61 & 12.00 & 411 &  47 & -80 &  45 & 4932.00 \\ 
  23 & 2001-01-10 & -3.69 & 5.30 & 367 &  13 &   0 &   5 & 1945.10 \\ 
  24 & 2000-05-16 & -3.71 & 8.40 & 444 &  30 &  -8 &  16 & 3729.60 \\ 
  25 & 2000-02-12 & -3.76 & 14.70 & 553 &  50 & -76 &  60 & 8129.10 \\ 
  26 & 2001-08-18 & -3.76 & 11.80 & 518 &  27 & -43 &  15 & 6112.40 \\  
27 & 2001-01-03 & -3.78 & 6.80 & 351 &  20 &  -8 &   8 & 2386.80 \\
28 & 2002-12-06 & -3.84 & 7.60 & 394 &  23 &  -6 &   9 & 2994.40 \\ 
  29 & 2000-09-26 & -3.89 & 5.90 & 566 &  37 & -41 &  24 & 3339.40 \\
  30 & 2000-08-31 & -3.93 & 5.00 & 546 &  27 & -18 &  14 & 2730.00 \\ 
  31 & 2000-09-29 & -4.01 & 5.50 & 378 &  17 & -19 &   7 & 2079.00 \\ 
  32 & 2002-04-18 & -4.06 & 6.60 & 613 &  27 & -24 &  15 & 4045.80 \\ 
  33 & 2002-09-19 & -4.06 & 12.80 & 485 &  53 & -104 &  63 & 6208.00 \\ 
  34 & 2002-05-28 & -4.11 & 5.60 & 662 &  23 & -26 &   9 & 3707.20 \\ 
  35 & 2000-10-01 & -4.16 & 4.30 & 418 &  27 & -37 &  13 & 1797.40 \\ 
  36 & 2000-09-03 & -4.18 & 6.80 & 413 &  17 & -14 &   7 & 2808.40 \\ 
  37 & 2000-12-11 & -4.20 & 4.00 & 552 &  20 &   0 &   8 & 2208.00 \\ 
  38 & 2000-05-08 & -4.21 & 9.80 & 360 &  10 &  19 &   4 & 3528.00 \\ 
  39 & 2000-10-05 & -4.33 & 13.40 & 486 &  67 & -138 & 116 & 6512.40 \\ 
  40 & 2000-04-07 & -4.36 & 9.90 & 573 &  50 & -162 &  74 & 5672.70 \\ 
  41 & 2000-05-30 & -4.38 & 8.30 & 445 &  30 & -37 &  18 & 3693.50 \\ 
  42 & 2001-10-09 & -4.38 & 6.20 & 617 &  37 & -29 &  22 & 3825.40 \\ 
  43 & 2002-07-18 & -4.38 & 5.70 & 441 &  13 &  -8 &   5 & 2513.70 \\ 
  44 & 2000-11-24 & -4.40 & 8.80 & 406 &  23 &   0 &  11 & 3572.80 \\ 
  45 & 2002-12-01 & -4.41 & 6.40 & 506 &  33 & -32 &  18 & 3238.40 \\ 
  46 & 2002-11-25 & -4.43 & 7.00 & 460 &  30 & -46 &  15 & 3220.00 \\ 
\hline
\end{tabular}
\end{table} 

\begin{table}[ht]
\label*{table 3}
\centering
\begin{tabular}{rlrrrrrrr}
  \hline
S/N & Date & FD(\%) & IMF (nT) & SWS (km$s^{-1}$) & kp & Dst (nT) & ap & SWSx IMF(km$s^{-1}$)(nT) \\ 
             
  \hline

  47 & 2002-05-20 & -4.45 & 9.00 & 453 &  23 & -12 &  10 & 4077.00 \\ 
  48 & 2000-08-26 & -4.46 & 6.00 & 385 &  13 &  -8 &   5 & 2310.00 \\ 
  49 & 2000-11-04 & -4.48 & 12.00 & 436 &  40 & -22 &  26 & 5232.00 \\ 
  50 & 2002-09-28 & -4.48 & 10.70 & 307 &  13 &   3 &   5 & 3284.90 \\ 
  51 & 2000-10-07 & -4.58 & 9.20 & 306 &  33 & -38 &  21 & 2815.20 \\ 
  52 & 2000-12-23 & -4.58 & 3.80 & 391 &  10 & -36 &   4 & 1485.80 \\  
53 & 2000-12-25 & -4.60 & 7.90 & 436 &  17 & -41 &   6 & 3444.40 \\
54 & 2001-04-19 & -4.60 & 11.90 & 352 &  17 &  10 &   6 & 4188.80 \\ 
  55 & 2002-10-01 & -4.62 & 19.50 & 388 &  50 & -100 &  67 & 7566.00 \\
56 & 2001-01-25 & -4.67 & 3.50 & 387 &  13 & -27 &   6 & 1354.50 \\ 
  57 & 2001-10-28 & -4.68 & 11.20 & 450 &  47 & -99 &  44 & 5040.00 \\ 
  58 & 2000-11-13 & -4.90 & 7.70 & 390 &  20 &  -1 &   8 & 3003.00 \\ 
  59 & 2000-12-27 & -4.90 & 4.50 & 578 &  23 & -18 &   9 & 2601.00 \\ 
  60 & 2000-11-16 & -4.92 & 4.80 & 391 &   7 &   2 &   3 & 1876.80 \\ 
  61 & 2002-12-08 & -5.02 & 7.10 & 599 &  27 & -28 &  12 & 4252.90 \\ 
  62 & 2002-11-03 & -5.07 & 9.70 & 478 &  43 & -65 &  35 & 4636.60 \\ 
  63 & 2002-02-01 & -5.09 & 11.20 & 347 &  27 & -17 &  14 & 3886.40 \\ 
  64 & 2002-04-24 & -5.10 & 7.10 & 488 &  17 & -30 &   7 & 3464.80 \\ 
  65 & 2002-03-22 & -5.14 & 7.50 & 444 &  13 &  -4 &   6 & 3330.00 \\ 
  66 & 2002-10-03 & -5.17 & 11.50 & 464 &  47 & -78 &  45 & 5336.00 \\ 
  67 & 2002-04-20 & -5.19 & 9.30 & 376 &   7 &  -8 &   2 & 3496.80 \\ 
  68 & 2002-05-23 & -5.19 & 17.00 & 606 &  47 & -38 &  78 & 10302.00 \\ 
  69 & 2002-09-24 & -5.19 & 10.10 & 563 &  53 & -106 &  70 & 5686.30 \\ 
  70 & 2002-07-23 & -5.20 & 4.80 & 472 &  30 & -12 &  17 & 2265.60 \\ 
  71 & 2001-04-01 & -5.22 & 9.60 & 458 &  37 & -61 &  26 & 4396.80 \\ 
  72 & 2001-10-22 & -5.22 & 7.50 & 746 &  43 & -137 &  38 & 5595.00 \\ 
  73 & 2002-09-11 & -5.22 & 15.10 & 578 &  60 & -150 &  96 & 8727.80 \\ 
  74 & 2001-04-05 & -5.25 & 7.50 & 617 &  33 & -31 &  19 & 4627.50 \\ 
  75 & 2002-01-30 & -5.27 & 5.90 & 338 &   7 &  -1 &   3 & 1994.20 \\ 
  76 & 2001-05-28 & -5.30 & 9.10 & 505 &  33 &  -8 &  18 & 4595.50 \\ 
  77 & 2002-10-25 & -5.37 & 6.80 & 689 &  43 & -68 &  39 & 4685.20 \\
78 & 2002-08-09 & -5.45 & 8.20 & 397 &  27 &  -7 &  14 & 3255.40 \\    
  79 & 2000-09-15 & -5.47 & 6.70 & 363 &  20 &   4 &  12 & 2432.10 \\ 
  80 & 1999-12-13 & -5.51 & 11.40 & 489 &  33 & -46 &  26 & 5574.60 \\ 
  81 & 2002-11-27 & -5.52 & 9.80 & 538 &  33 & -50 &  24 & 5272.40 \\ 
  82 & 2001-04-16 & -5.57 & 4.00 & 453 &  20 & -24 &   8 & 1812.00 \\ 
  83 & 2000-09-09 & -5.61 & 8.90 & 502 &  20 & -28 &   8 & 4467.80 \\ 
  84 & 2002-09-04 & -5.61 & 4.40 & 425 &  13 &  -9 &   5 & 1870.00 \\ 
  85 & 2002-12-15 & -5.61 & 12.60 & 422 &  47 & -72 &  42 & 5317.20 \\ 
  86 & 2001-10-12 & -5.76 & 11.40 & 501 &  40 & -51 &  34 & 5711.40 \\ 
  87 & 2000-07-11 & -5.86 & 13.70 & 458 &  43 &  13 &  34 & 6274.60 \\ 
  88 & 2000-06-20 & -5.94 & 6.20 & 379 &  17 &   5 &   6 & 2349.80 \\ 
  89 & 2000-06-24 & -6.01 & 9.00 & 551 &  27 & -20 &  15 & 4959.00 \\ 
  90 & 2002-08-07 & -6.01 & 4.70 & 343 &  10 &   2 &   4 & 1612.10 \\ 
  91 & 2002-09-08 & -6.01 & 11.70 & 479 &  33 & -101 &  36 & 5604.30 \\ 
  92 & 2002-10-21 & -6.03 & 5.80 & 571 &  20 & -18 &   8 & 3311.80 \\ 
  93 & 2000-10-29 & -6.09 & 13.70 & 381 &  40 & -89 &  34 & 5219.70 \\ 
  94 & 2002-07-20 & -6.13 & 7.40 & 789 &  30 & -20 &  18 & 5838.60 \\ 
  95 & 2000-06-26 & -6.19 & 5.80 & 343 &  17 &  -1 &   7 & 1989.40 \\ 
\hline
\end{tabular}
\end{table} 

\begin{table}[ht]
\label*{table 3}
\centering
\begin{tabular}{rlrrrrrrr}
  \hline
S/N & Date & FD(\%) & IMF (nT) & SWS (km$s^{-1}$) & kp & Dst (nT) & ap & SWSx IMF(km$s^{-1}$)(nT) \\ 
             
  \hline

  96 & 2000-12-06 & -6.19 & 11.50 & 512 &  43 & -36 &  40 & 5888.00 \\ 
  97 & 2000-11-11 & -6.21 & 8.40 & 545 &  37 & -46 &  24 & 4578.00 \\ 
  98 & 2002-11-05 & -6.21 & 7.20 & 804 &  30 & -35 &  16 & 5788.80 \\ 
  99 & 2002-03-25 & -6.29 & 15.40 & 433 &  17 & -33 &   7 & 6668.20 \\ 
  100 & 2001-04-09 & -6.43 & 8.60 & 622 &  33 & -53 &  20 & 5349.20 \\ 
  101 & 2002-08-23 & -6.60 & 8.80 & 402 &  17 & -18 &   7 & 3537.60 \\ 
  102 & 2002-01-11 & -6.63 & 8.90 & 610 &  40 & -42 &  27 & 5429.00 \\ 
  103 & 2002-12-20 & -6.70 & 6.10 & 528 &  33 & -47 &  21 & 3220.80 \\ 
  104 & 2002-08-20 & -6.71 & 7.20 & 479 &  33 & -48 &  30 & 3448.80 \\ 
105 & 2001-11-07 & -6.75 & 6.50 & 635 &  30 & -110 &  19 & 4127.50 \\
  106 & 2001-04-29 & -6.92 & 7.60 & 596 &  23 & -18 &  13 & 4529.60 \\ 
  107 & 2001-08-29 & -7.02 & 4.20 & 459 &  13 &  -7 &   5 & 1927.80 \\ 
  108 & 2002-08-28 & -7.65 & 8.90 & 447 &  17 & -19 &   7 & 3978.30 \\ 
  109 & 2002-11-12 & -7.13 & 12.40 & 569 &  30 & -15 &  17 & 7055.60 \\ 
  110 & 2002-01-03 & -7.29 & 5.90 & 342 &   7 & -16 &   2 & 2017.80 \\ 
  111 & 2002-12-23 & -7.47 & 10.10 & 517 &  37 & -42 &  24 & 5221.70 \\ 
  112 & 2000-11-07 & -7.79 & 20.20 & 512 &  43 & -89 &  46 & 10342.40 \\ 
  113 & 2001-10-01 & -7.81 & 11.10 & 498 &  47 & -99 &  48 & 5527.80 \\ 
  114 & 2000-05-24 & -7.87 & 13.70 & 636 &  60 & -90 &  93 & 8713.20 \\ 
  115 & 2000-07-29 & -7.92 & 8.80 & 460 &  37 & -38 &  27 & 4048.00 \\ 
  116 & 2000-07-29 & -7.92 & 10.70 & 549 &  33 & -72 &  26 & 5874.30 \\ 
  117 & 2000-08-06 & -7.97 & 6.00 & 515 &  30 & -32 &  16 & 3090.00 \\ 
  118 & 2000-07-13 & -8.33 & 11.10 & 573 &  37 &  -4 &  42 & 6360.30 \\ 
  119 & 2001-11-25 & -8.43 & 11.50 & 650 &  20 & -106 &   8 & 7475.00 \\ 
  120 & 2000-08-12 & -8.86 & 25.00 & 599 &  67 & -128 & 123 & 14975.00 \\ 
  121 & 2002-11-18 & -9.02 & 9.30 & 378 &  23 & -37 &  10 & 3515.40 \\ 
  122 & 2002-08-02 & -9.12 & 12.10 & 489 &  43 & -59 &  42 & 5916.90 \\ 
  123 & 2000-09-18 & -9.59 & 19.20 & 744 &  53 & -103 &  70 & 14284.80 \\ 
  124 & 2000-06-09 & -9.69 & 10.20 & 609 &  13 & -34 &   5 & 6211.80 \\ 
  125 & 2000-11-29 & -9.77 & 9.20 & 512 &  47 & -81 &  56 & 4710.40 \\ 
  126 & 2000-07-22 & -10.21 & 5.70 & 425 &  27 & -40 &  12 & 2422.50 \\ 
  127 & 2000-07-20 & -10.88 & 8.10 & 533 &  43 & -67 &  36 & 4317.30 \\ 
  128 & 2001-04-12 & -12.79 & 15.10 & 659 &  40 & -131 &  50 & 9950.90 \\ 
  129 & 2000-07-16 & -15.58 & 21.80 & 816 &  43 & -172 &  50 & 17788.80 \\ 
   \hline
\end{tabular}
\end{table}

\section {Acknowledgments}

It is a pleasure to thank the team that hosts the websites {http://cr0.izmiran
.rssi.ru/ and https://omniweb.gsfc.nasa.gov/html/ow data.html} from where we sourced the data for this article. We also appreciate the kind contributions of the ananymous reviewer. We want to acknowledge in a sepcial way, the non-commercial R software  developers and all our friends on the  R-mailing list (R-help@r-project.org), especially Rui Barradas, for his assistance during  the  preliminary data processing stage. To you all, we are sincerely grateful. 
  

\bibliography{reference}

\end{document}